\documentclass[12pt]{article}
\usepackage{epsf,epsfig}
\def \lta {\mathrel{\vcenter{\hbox{$<$}\nointerlineskip\hbox{$\sim$}}}}
\def \gta {\mathrel{\vcenter{\hbox{$>$}\nointerlineskip\hbox{$\sim$}}}}

\begin{document}

\begin{titlepage}
\pagestyle{empty}
\baselineskip=21pt

\begin{center}
{\large{\bf Moduli from Cosmic Strings}}
\end{center}
\begin{center}
\vskip 0.05in
{{\bf Marco Peloso}$^1$ and {\bf Lorenzo Sorbo}$^2$
\vskip 0.05in
{\it
$^1${Physikalisches Institut, Universit{\"{a}}t Bonn,\\
Nussallee 12, D-53115 Bonn, Germany}\\
$^2${Institut d'Astrophysique de Paris, C.N.R.S.,\\ 98bis Boulevard Arago, 75014 Paris, France}\\
}}
\vskip 0.35in
{\bf Abstract}
\end{center}
\baselineskip=18pt \noindent
Moduli fields are generally present in superstring--inspired models. They are
typically characterized by masses of the order of the supersymmetry breaking
scale and by interactions of gravitational strength to ordinary matter. If stable, they can easily overclose the Universe. If unstable, strong limits on their abundance have to be imposed not to disrupt the
successful predictions of nucleosynthesis. We discuss the production of moduli
quanta from loops of cosmic strings, which may form at phase transitions in the
early history of the Universe. As an application, we then focus on strings
formed at the end of hybrid inflation, showing that in this case the coupling
of the moduli to the field which forms the string network has to be much
weaker than the typical gravitational interactions not to exceed the bounds
from nucleosynthesis. Alternatively, if the relevant coupling is of gravitational order, an upper bound of about $10^{13}$~GeV is imposed on the string energy scale.
\vfill
\vskip 0.15in
\leftline{May 2002}
\end{titlepage}
\baselineskip=18pt

\section{Introduction}

Most superstring--inspired models of particle physics predict the
existence of moduli fields, that are generally characterized by a mass in
the TeV range (that is, of the order of the expected scale of
supersymmetry breaking) and by gravitationally suppressed couplings to
ordinary matter. These properties are shared by their supersymmetric
partners, the so called modulinos, and by the partners of gravitons, the
gravitinos, which are also present in these models. Due to their extremely
weak couplings, such particles are not likely to be observed in the
foreseeable accelerator experiments. However, they can play a relevant role
in cosmology~\cite{wei}.

In standard cosmology, particles which interact only gravitationally with the
thermal background are often denoted altogether as gravitational relics. They
have a very early decoupling, and, as a consequence, they can be expected to
have a significant abundance. If stable, gravitational relics with a mass above
the keV range lead to the overclosure of the Universe. If unstable, they
disrupt the successful predictions of big bang nucleosynthesis, unless their
mass is above $20\,$ TeV, so that nucleosynthesis can occur after they have
decayed~\cite{wei}.

These limits do not hold in inflationary scenarios~\cite{infla}, where the
primordial abundance (if any) is diluted away during inflation. However, in
this case one has to check that gravitational relics are not (re)generated to a
too high abundance after inflation~\cite{reh}. For unstable relics $s$ with
lifetimes $> 10^4\,$ sec, the most restrictive bound comes from
photo-destruction of the light elements produced during
nucleosynthesis~\cite{grnuc}
\begin{equation}
Y_s  \lta 10^{-14} \left( {\rm TeV} / m_s \right) \,\,.
\label{bound}
\end{equation}

The strength of this bound has led to investigate possible overproduction
of gravitational relics from several different sources. The most standard
studies concern the production from the thermal bath formed at reheating.
It has been shown~\cite{infla,reh,grnuc} (see also~\cite{boundtr} for more
recent discussions) that the bound~(\ref{bound}) translates into the upper
limit $T_{\rm rh} \lta 10^9 \, {\rm GeV}\,$ on the reheating temperature.
More recently, attention has been drawn to nonthermal production of gravitinos~\cite{grapreh} and moduli~\cite{gramod} during the coherent oscillations of the inflaton field (preheating), while production of gravitinos from inflaton/inflatino decays has been discussed in~\cite{nop}.

In the present work we consider topological defects, and in particular cosmic
strings, as possible source of gravitational relics. For a long time cosmic
strings (for a review, see~\cite{kib}) have represented a viable alternative to
inflation in the explanation of the inhomogeneities we observe today in the
Universe~\cite{vile}. The recent accurate measurement of the CMB power
spectrum, however, seems to favor inflation as the main origin of primordial
density perturbations. Nevertheless, the presence of a subdominant component of
matter in the Universe in the form of a network of cosmic strings is not
excluded~\cite{durrer}. Moreover, one of the most studied inflationary
scenarios, hybrid inflation~\cite{hybrid}, predicts the formation of a network
of cosmic strings at the beginning of reheating. Motivated by this observation,
although the formalism we will use is fairly general, in the following analysis
we will mainly refer to a cosmic string network that formed at the end of
hybrid inflation, with a symmetry breaking scale of the order of
$10^{15}\,$GeV.

Cosmic strings can lead to particle production. Actually, they have already
been considered as potential sources of ultra high energy cosmic
rays~\cite{bona} or of gamma--ray bursts~\cite{bere}, as well as a possible
origin of nonthermal cold dark matter~\cite{jzb}, or as source of the
matter-antimatter asymmetry~\cite{bario}. What is more relevant for the present
discussion, it has also been suggested in reference~\cite{kklv} that string
loops may lead to a significant production of gravitinos.  Moreover, in
ref.~\cite{davi}, bounds on the cosmic string scale have been obtained by
analyzing string coupling to the dilaton of (super)string theory. 

Here, we will consider loops of cosmic strings as possible sources of scalar
moduli. Moduli degrees of freedom in general parameterize symmetries in
superstring models. These symmetries are usually broken by supersymmetry
breaking effects, so that moduli acquire a mass of the order of the
supersymmetry breaking scale. In superstring--inspired models, the expectation
value of the moduli fields determines the coupling constants and the masses of
the theory. By expanding such coupling constants in powers of the moduli
fluctuations, it is possible to see that matter fields can be generally
expected to have a Planck mass suppressed coupling to moduli. In the present
work, we assume that this is the case also for the field(s) responsible for the
formation of cosmic strings.

A simple example related to the application we are discussing is given by the following simple (toy) model with a $D$--term
\begin{equation}
V \supset - \, \frac{1}{2\,f^2} \, D^2 + D \left( \phi^+ q \phi - \xi \right) \,\,,
\label{vd}
\end{equation}
where $\phi^T = \left( \phi_+ \,,\, \phi_- \right)$ contains two scalar fields with opposite $U \left( 1 \right)$ charges, $q = {\rm diag } \, \left( + 1 \,,\, - 1 \right) \,\,$, and $\xi$ is a Fayet-Iliopoulos term. The coupling constant $f$ of the gauge kinetic term is determined by the expectation value of a scalar modulus $\tilde{s}$. Defining $s\equiv \tilde{s}-\langle \tilde{s} \rangle$, we can expand
\begin{equation}
f = f_0 \left( \langle \tilde{s} \rangle  \right) \, \left( 1 + \frac{\alpha}{8} \, \frac{{\rm Re }\, s}{M_P} + \dots \right)
\label{dila}
\end{equation}
(here and in the following, dots denote higher order terms in $s/M_P$ which we
neglect; moreover, we assume $f_0$ and $\alpha$ to be real). Hereafter,
${\rm Re } \, s \equiv s\,$. Eliminating $D\,$ in eq. (\ref{vd}), and
making use of the expansion (\ref{dila}), we get
\begin{equation}\label{seriess}
V \supset \frac{1}{2} \, f_0^2 \left( 1 +  \frac{\alpha}{4} \, \frac{s}{M_P} + \dots \right) \left( \vert \phi_+ \vert^2 - \vert \phi_- \vert^2 - \xi \right)^2
\,\,.
\end{equation}
We can assume the vacuum expectation value of $\phi_-$ to vanish. Finally,
renaming $f_0^2 \equiv \lambda/2 \;,\; \phi_+ \equiv \phi \;,\; \xi =
\eta^2 \,$, the potential term is recast in the form
\begin{eqnarray}\label{pot}
V &=& \frac{\lambda}{4} \left( \vert \phi \vert^2 - \eta^2 \right) + \frac{\lambda}{4}\,\alpha\,\frac{s}{M_P} \,\left( \vert \phi \vert^2 -\eta^2\right)^2+ \dots \simeq\nonumber \\
&&\simeq \frac{\lambda}{4} \left( \vert \phi \vert^2 - \eta^2 \right) + \lambda\,\alpha\,s\frac{\eta^2}{M_P} \,\left( \vert \phi \vert -\eta\right)^2+ \dots
\end{eqnarray}

The first term in the above potential leads to the spontaneous breaking of the
$U\left(1\right)$ symmetry $\phi \rightarrow e^{i\,\theta}\,\phi$. The second
term describes a cubic interaction between $s$ and $\phi$. We assume that in
the early Universe the $U\left(1\right)$ symmetry is restored either due to
finite temperature corrections to the potential or nonthermally. This second
possibility is naturally implemented in hybrid inflation, where the field
responsible for the $U \left( 1 \right)$ breaking is coupled to the inflaton
$\psi\,$, and the symmetry gets restored at high values of $\psi\,$.

When the symmetry breaks, a string network is expected to form. Eventually,
string loops are generated from the network. The oscillations of these loops
act as a source of quanta of the modulus field. The rate of quanta emitted by a
single loop has been estimated by Srednicki and Theisen~\cite{srth}, and we
will  briefly review this calculation in section~\ref{unaloop}. This
computation applies if one can neglect the coupling between the loop and the
thermal background. This turns out to be the case only at sufficiently late
times, when the string network enters the so called scaling regime (see below
for details). In section~\ref{tanteloop} we then compute the total number
density of quanta produced in this regime. Although we neglect the possible
production at earlier times, we will see that the bound~(\ref{bound}) can
impose a significant upper limit on the coupling $\alpha$ between the string
and the modulus field. In section~\ref{scale}, on the other hand, we assume
$\alpha\sim 1$, such that an upper bound on $\eta$ is derived, by also taking
into account effects related to cusp annihilations. We compare this bound with
the corresponding one given in ref.~\cite{davi}. Our conclusions are finally
presented in section~\ref{concl}.

\section{Rate of quanta emission by an oscillating loop}\label{unaloop}

In this section we perform the calculation of the rate of emission of quanta of the scalar $s$ by an oscillating Nambu--Goto string loop. The interaction terms we consider are given in eq.~(\ref{pot}). The calculation is analogous to the one presented in ref.~\cite{srth} for the case of the coupling $s^2\,\left\vert \phi\right\vert ^2\,$.

We always consider a regime in which the string thickness can be neglected. In this approximation, a reparametrization invariant expression for the expectation value of $(|\phi|-\eta)^n$ in the string state $\vert S\rangle$ is given by
\begin{eqnarray}\label{repinv}
\langle S \vert\, \left(|\phi\left(x^\mu\right)|-\eta\right)^n\,\vert S \rangle\simeq
\eta^n\,a^2\,\int_0^L d\sigma\,\int d\tau \sqrt{-{\rm {det}}\,g}\,\delta^{\left(4\right)}\left(x^\mu-s^\mu\left(\sigma,\,\tau\right)\right),\nonumber\\
\end{eqnarray}
where $a\simeq \left( \eta \, \sqrt{\lambda} \right)^{-1}$ is the string thickness, $\sigma\,$ and $\tau$ are coordinates on the string worldsheet, and $s^\mu$ denotes the location of the core of the string. Since we are considering a closed string loop, $\sigma$ ranges from $0$ to $L$. Finally, ${\rm {det}}\,g$ is the determinant of the induced metric on the string worldsheet.

The modulus $s$, of mass $m_s$, is coupled to the string via the trilinear
interaction~(\ref{pot}).
 As a consequence, the amplitude of transition
from a string state $|S\rangle$ to a state
$|S',\,s\left(k^\mu\right)\rangle$ containing the string and a quantum of
$s$ of momentum $k^\mu$ is given, via LSZ reduction, by
\begin{eqnarray}\label{ampli}
\langle S',\,s\left(k^\mu\right)|S\rangle &&\simeq \int d^4x\, e^{i\,k^\mu x_\mu}\,\langle S'|\left(\partial_\mu\,\partial^\mu + m_s^2\right)\,s\left(x^\mu\right)|S\rangle\simeq\nonumber\\
&&\simeq \lambda\,\alpha\,\frac{\eta^2}{M_P}\,\int d^4x\, e^{i\,k^\mu x_\mu}\,\langle S'|\left(|\phi|-\eta\right)^2|S\rangle\,\,.
\end{eqnarray}

To proceed, we assume that $|S'\rangle\simeq |S\rangle$ (that is, that the emission of a quantum of $S$ does not alter substantially the string wave function). Then, in eq.~(\ref{repinv}), we choose a frame in which $\tau=t=s^0(\sigma,\,t)$, such that, after integration over $d^3x$, we obtain
\begin{equation}
\langle S',\,s\left(k^\mu\right)|S\rangle\simeq \alpha\,\frac{\eta^2}{M_P}\,\int_{-\infty}^{+\infty} dt\,e^{iEt}\,\int_0^L d\sigma\, \vert \vec{s}\,{}'\left(\sigma,\,t\right) \vert^2\,e^{-i\vec{k}\vec{s}\left(\sigma,\,t\right)}\,\,,
\end{equation}
where $E=k^0$ is the energy of the quantum of $s$ emitted, and $\vec{k}$ its momentum. We denote with a prime the derivative with respect to $\sigma$ and with a dot the derivative with respect to $t$.

The dynamical equations for the Nambu--Goto string read
\begin{equation}
\ddot{\vec{s}}-\vec{s}\,{}''=0\,,\qquad \left(\dot{\vec{s}}\pm \vec{s}\,{}'\right)^2=1\,,
\label{streq}
\end{equation}
and the periodicity of the loop requires $\vec{s}\left(\sigma,\,t\right) = \vec{s}\left(\sigma+L,\,t\right)$. From eqs.~(\ref{streq}), one can also show~\cite{kib} that $\vec{s} \left( \sigma + L/2 ,\,t + L/2 \right) = \vec{s}\left(\sigma,\,t\right)\,$, so that the motion of the string loop has actually period $L/2\,$. Hence, we can decompose
\begin{equation}
\int_0^L d\sigma\, \vert \vec{s}\,{}'\left(\sigma,\,t\right) \vert^2\,e^{-i\vec{k}\vec{s}\left(\sigma,\,t\right)}\equiv\sum_n \, a_n \left( \vec{k} \right) \,e^{-4\pi i n t/L}\,\,.
\label{deco}
\end{equation}
As a consequence, 
the power emitted by the string into quanta of $s$ can be expressed as
\begin{equation}
\frac{dE_s}{dt}\,T=\int \frac{d^3k}{\left(2 \pi \right)^3\,2E}\,E\,\left|\langle S',\,s\left(k^\mu\right)|S\rangle \right|^2=T\,\sum_{n=0}^\infty P_n\,\,,
\end{equation}
where $T=2\pi\,\delta\left(0\right)$ is the total duration of the process, and
\begin{equation}\label{pienne}
P_n=2\pi\,\left(\alpha\,\frac{\eta^2}{M_P}\right)^2\,\int\frac{d^3k}{\left(2 \pi \right)^3\,2E}\,E\,\left|a_n\right|^2\,\delta\left(E-4\pi n/L\right)\,\,.
\end{equation}

The coefficients $a_n$
\begin{equation}\label{aenne}
a_n\equiv\frac{2}{L}\int_0^{L/2}dt\,\int_0^L d\sigma\, e^{iEt-i\vec{k}\vec{s}}\, \vert \vec{s}\,{}'\left(\sigma,\,t\right) \vert^2\,\,
\end{equation}
can be estimated via the stationary phase method~\cite{srth}, which
consists in integrating the function $\vert \vec{s} \, {}' \left( \sigma ,
\, t \right) \vert^2$ only over the regions in which the phase $\psi \equiv
E \, t -  \vec{k} \, \vec{s}$ is nearly constant. Out of these regions,
the phase $\psi$ rapidly oscillates, so that the integral there averages to
negligible values. Requiring $\psi' = 0$ forces $\vec{s}\,{}'=0\,$, which
in turn implies $\vert \dot{\vec{s}} \vert = 1\,$. Hence, at these points
the string moves at the  speed of light and, provided  $\vec{s}\,{}''$ is
nonvanishing, takes the appearance of a cusp.

Let us consider one of these points and choose coordinates such that
$t=\sigma=0,\, \vec{s} = \vec{0}$ there. For definiteness, the region of
nearly constant phase will be considered the one for which $\vert \psi
\vert < 1\,$. By expanding $\psi$ at small $t$ and $\sigma$, it is
possible to show~\cite{srth} that this region extends up to $\vert \sigma
\vert ,\, \vert t \vert \lta \sigma_{\rm {max}} = L \left( L \, E \right)^{-\,1/3}\,$, and that
the coefficients $a_n$ can be estimated to be $a_n \sim L \, \left( L \, E
\right)^{-\,4/3}\,$. One can also show~\cite{srth} that these regions are
actually nonvanishing only provided that $\vec{k}$ lies in a cone centered
around $\dot{\vec{s}}$ and of amplitude $\epsilon\simeq
\left(L\,E\right)^{-1/3}$, and provided that $E > E_{\rm min} \equiv
m_s^{3/2} \, L^{1/2}\,$.

From these results, the coefficients $P_n$ can be (somewhat crudely,
neglecting all the numerical coefficients) estimated to be
\begin{eqnarray}\label{pienne1}
P_n &\sim& \alpha^2\,\left( \frac{\eta^2}{M_P} \right)^2\int k^2\,dk\,d\Omega\,\left|a_n\right|^2 \, \delta \left(E-4\pi\,n/L\right) \nonumber\\
&\sim& \alpha^2 \, \left( \frac{\eta^2}{M_P} \right)^2 \,E^2 \,
\epsilon^2 \,\left|a_n\right|^2 \sim \alpha^2\,\left(\frac{\eta^2}{M_P}\right)^2 \, \frac{1}{n^{4/3}} \,\,,
\end{eqnarray}
where the behaviour $P_n\propto n^{-4/3}$ is in agreement with the analogous expression in~\cite{davi}.

The total power emitted by the string is obtained by summing over the modes $n\,$. We distinguish two cases. For $L > m_s^{-1}$, the sum starts from $n \simeq E_{\rm min} \, L \simeq \left( m_s L \right)^{3/2} \,$ (notice that we have $E \geq E_{\rm min} > m_S\,$ in this case). In the opposite case, $L \leq m_s^{-1}\,$, one simply starts from $n=1$ (which also ensures $E \geq L^{-\,1} > m_s$). By approximating the sum over $n$ with an integral, we get\footnote{Notice that, for $L\lta 1/m_s\,$, the following expression agrees, up to numerical factors, with the power emitted into gravitational waves $P_g$ (see below). Indeed, this is expected, since this regime corresponds to emission of one massless particle, as in the case of graviton production~\cite{kib}.}
\begin{equation}\label{dedt}
P_s\equiv \frac{dE_s}{dt}\sim \left\{ \begin{array}{l} \alpha^2\,\left(\frac{\eta^2}{M_P}\right)^2\,\frac{1}{\sqrt{m_s\,L}}\,\,,\qquad L\gg 1/m_s\,\,,\\
\alpha^2\,\left(\frac{\eta^2}{M_P}\right)^2\,\,,\qquad L\lta 1/m_s\,\,.
\end{array}\right.
\end{equation}

In an analogous way, it is possible to estimate the number of quanta of $s$ emitted per unit time by the string oscillations
\begin{equation}\label{dndt}
\frac{dN_s}{dt}\sim\left\{ \begin{array}{l} \alpha^2\,\left(\frac{\eta^2}{M_P}\right)^2\,\frac{1}{m_s}\,\frac{1}{m_s\,L}\,\,,\qquad L\gg 1/m_s\,\,,\\
\alpha^2\,\left(\frac{\eta^2}{M_P}\right)^2\,\frac{1}{m_s}\,\left(m_s\,L\right)\,\,,\qquad L\lta 1/m_s\,\,.
\end{array}\right.
\end{equation}

To compute the total number of quanta of $s$ emitted, this last expression must
be integrated over the life of the loop. We do so for a regime in which the
interaction of the loop with the thermal background can be neglected (that is,
after the end of the so called ``friction dominated regime'', see section
\ref{tanteloop} for more details). In this case, the evolution of the loops is
determined only by their tension and by their decays.

Besides the production of quanta of $s$, the loop decays by emission of
gravitons, with energy per unit time $P_g = \Gamma \, \eta^4 / M_P^2 \,$, where
$\Gamma$ is a numerical factor (of the order of $50-100$) dependent on the
shape of the loop~\cite{kib}. Moreover, we consider the possibility that $\phi$
is coupled with some other scalar field $w$ (representing for example an MSSM
field) via the interaction term $g \, \phi^2 \, w^2\,$. From this quartic
coupling, the energy per unit time of quanta of $w$ produced by the loop is
$P_w = \left(g^2/\lambda^{3/2}\right) \eta / L \,$~\cite{srth}. The decay into
quanta of $s\,$ is always subdominant with respect to the one into gravitons,
provided that $\alpha^2/\Gamma \ll 1$ (in the following, this relation will be
always assumed to hold). By comparing $P_g$ and $P_w\,$, one instead sees that
the loop mainly decays into gravitons as long as its radius satisfies
\begin{equation}\label{lbar}
L > {\bar L} \equiv \frac{g^2}{\lambda^{3/2}}\,\frac{1}{\Gamma} \, \frac{M_P^2}{\eta^3} \simeq 10^{-\,6} \, {\rm TeV}^{-\,1} \, \frac{g^2}{\lambda^{3/2}}\,\left(\frac{100}{\Gamma}\right) \left( \frac{10^{15} \, {\rm GeV}}{\eta} \right)^3 \,\,,
\end{equation}
while the decay into quanta of $w$ dominates for smaller lengths. We conclude that, as long as $L\gta \bar{L}$, the radius of the loop reduces according to
\begin{equation}\label{radius}
L \left( t \right) \simeq L_F - \Gamma \, \frac{\eta^2}{M_P^2} \, \left( t - t_F \right)\;\;,\;\; {\bar L} < L < L_F\,\,,
\end{equation}
where $L_F = L \left( t_F \right)$ is the radius of the loop when it forms. From here on, we will assume that loops form with $L_F\gta \bar{L}\,$ since this will be the case in the main application we will consider in the next section. More precisely, since we will be interested in the production of quanta of $s$ with mass at the TeV scale, we assume here the hierarchy $L_F > m_s^{-\,1} > {\bar L}\,$, the latter inequality being satisfied as long as $\eta$ is not much smaller than $10^{15}\,$GeV, see eq.~(\ref{lbar}).

By combining eqs.~(\ref{dndt}) and (\ref{radius}), one sees that the quanta of $s$ are mainly produced when $L \sim m_s^{-1} \,$. For $L_F\gg m_s^{-1}$, their total number is approximatively given by
\begin{equation}
N_s \left( L_F \right) \simeq \alpha^2 \, \frac{\eta^2}{\Gamma \, m_s^2} \, {\rm ln } \, \left( L_F\,m_s \, \right) \,\,.
\label{n1}
\end{equation}

It is worth remarking that -- even if we have assumed a gravitationally
suppressed coupling between the string and the field $s\,$, see eq.~(\ref{pot})
-- the final amount $N_s$ of moduli emitted by a string loop is not Planck mass
suppressed. This is due to the fact that, in the regime we are considering, 
the loop decays gravitationally, so its lifetime becomes infinite in the limit
$M_P \rightarrow \infty\,$ \cite{kklv}.

In ref.~\cite{brand} it has been noted that the results of~\cite{srth} may be
affected by nonperturbative processes leading to the annihilation of the cusps.
Indeed, it was observed~\cite{brand} that different portions of the loop in a
region centered at the cusp and characterized by  $\vert \sigma \vert \leq
\sigma_c \equiv \eta^{-1/3} \, L^{2/3}$ have a mutual distance smaller than the
width of the string. Thus, the approximation of a delta--like string breaks
down in this region, and the simple Nambu--Goto dynamics cannot be trusted.
This effectively puts an ultraviolet cut-off on the decomposition~(\ref{deco}).
To investigate the validity of the results~(\ref{dedt}) and (\ref{dndt}), we
have to ensure that the harmonics which dominate the production of the quanta
of $s$ are below this cut-off, or, in other words, that the part of the loop
responsible for the production is much longer than the part where
nonperturbative phenomena are expected to occur. This is the case for
$\sigma_{\rm max} \gg \sigma_c\,$, which is in turn satisfied provided that the
typical energy $E$ of the quanta at their emission is much smaller than the
scale of the $U \left( 1 \right)$ symmetry breaking $\eta\,$. As we have seen
in this section, the quanta of $s$ are mainly produced when the radius of the
loop has the size of their inverse mass, $L \sim m_s^{-\,1}\,$. We have also
seen that the typical energy of the quanta emitted at that time is $E \sim
m_s^{3/2} \, L^{1/2} \sim m_s$, which is indeed much smaller than $\eta\,$. We
remark that in the work~\cite{blpi} it has been shown that the Lorentz contraction of the thickness of the string reduces the portion of the loop where the overlapping occurs to $\sigma_c \sim \eta^{-1/2} \, L^{1/2}\,$. Thus, {\it a
fortiori}, $\sigma_{\rm max} \gg \sigma_c$ also in this case.

This analysis confirms the validity of the results~(\ref{dedt}) and (\ref{dndt}), also when nonperturbative effects are taken into account. The reason is that we have considered production of single quanta of $s$ via the trilinear coupling~(\ref{pot}). We have seen that this production is dominated by the lowest harmonics $n$ in the sum~(\ref{deco}). As we have discussed before eq.~(\ref{dedt}), for radii $L \sim m_s^{-\,1}$ the main production comes indeed from collective oscillations of the whole loop. The production of two quanta from a quadrilinear coupling $\phi^2 s^2\,$ which was studied in~\cite{srth} is instead dominated by the highest possible harmonics. In this case nonperturbative effects at small scales might be relevant~\cite{brand}.

To conclude this section, it is also useful to compute the total energy emitted into quanta of $w\,$. Assuming that these particles rapidly thermalize, it may (at least in principle) give a sizeable contribution to the entropy of the Universe. From the given $P_g$ and $P_w\,$, we find
\begin{equation}
E_w \simeq \frac{g^2\,M_P^2}{\lambda^{3/2} \Gamma \, \eta} \, {\rm ln } \, \left( \frac{L_F}{{\bar L}} \right) \,\,.
\label{ew}
\end{equation}
We will see in the next section that, in the range of parameters we are considering, these quanta will give a negligible contribution to the background energy density and entropy of the Universe.

\section{Quanta produced in the scaling dominated regime} \label{tanteloop}

Equation (\ref{n1}) gives an order of magnitude estimate of the number of moduli emitted by a loop that obeys the Nambu-Goto dynamics. In order to analyse the cosmological effects of the moduli that have been produced, we now need to estimate the number density of the loops.

The evolution of a string network can roughly be divided into two stages.
In the initial, friction dominated stage, the interaction of every string
segment with the background plasma cannot be neglected. In this regime,
the frictional drag of the background plasma smoothens out all structures
on lengthscales below $L_{\mathrm {min}}\sim \eta\, M_P^{1/2}/T^{5/2}\,$
\cite{kib}. The dynamics of the loops cannot be approximated as the one of
Nambu-Goto strings, and therefore the analysis that led to eq.~(\ref{n1})
cannot be trusted. Even if we expect that also in this epoch the decay of
string loops will lead to production of moduli, we will make the
conservative assumption of neglecting the moduli produced during the
friction dominated regime.

The friction dominated stage comes to an end when the scale $L_{\mathrm
{min}}$ becomes comparable with the Hubble length, that is, when the
temperature of the universe drops below $T_{\mathrm {fr}}\sim
\eta^2/M_P\,$ \cite{kib}. In the subsequent regime the dynamics
of loops smaller than the Hubble radius is well approximated by the
Nambu-Goto dynamics. The network in then expected to reach a scaling regime, in which the energy in strings is a constant fraction of the background energy in the Universe~\cite{kib}.

The system we are considering consists of cosmic strings formed at the end of the inflationary phase, with a mass per unit length of about $\eta^2\simeq \left( 10^{15} \, {\rm {GeV}} \right)^2$. For those strings, the temperature at which the scaling regime begins is of the order of $\eta^2/M_P \simeq 10^{12}\,{\rm {GeV}}$. However, due to thermal production of moduli, the reheating temperature is constrained to be smaller than about $10^9$~GeV. As a consequence, when radiation domination occurs, the string network is already in a regime in which the dynamics of the single loop is the Nambu--Goto dynamics.

For a string network evolved in a radiation dominated Universe, the number density of loops of strings created per unit time during the scaling regime is given by~\cite{ack}
\begin{equation}\label{numloops}
\frac{d n_L}{d t} = \frac{\nu}{\left(\kappa-1\right)\,\Gamma\,\left(\eta^2/M_P^2\right)\,t^4} \,\,,
\end{equation}
where $\nu$ is a constant of order one and $\kappa$ is a constant in the range $2<\kappa<10$.\footnote{Different estimates can be found in the literature (see e.g.~\cite{numdiv}) for the quantity $dn_L/dt$, with higher powers of $\eta/M_P$ at the denominator. Since this would enhance the production of quanta of $s\,$, in the following analysis we will conservatively assume the validity of eq.~(\ref{numloops}).}

In this regime, the typical length of a loop formed at the time $t$ amounts to~\cite{ack}
\begin{equation}\label{lenloops}
L_F\left(t\right) \simeq \left(\kappa-1\right)\,\Gamma\,\frac{\eta^2}{M_P^2}\, t\,\,.
\end{equation}

We have therefore all the elements to compute the total number of moduli
produced by the string network. Notice, however, that the above formulae
rely on the hypothesis that the string network evolved in a
radiation--dominated, thermalized background. Actually, this is not the
case for a network formed at the end of hybrid inflation, that
evolved during the reheating period. Nevertheless, it is likely that a
string network will be present after the end of reheating~\cite{jeann}. In order to
be able to estimate the abundance of moduli produced by the network, we will
make the assumption that also in our case eqs.~(\ref{numloops})
and~(\ref{lenloops}) become valid, once a thermal bath is formed and the
radiation dominated period starts. Moreover, since the dynamics of the
loops and of the network during reheating is unknown, we will
conservatively neglect all the processes that could have led to production
of quanta of $s$ during this period.

Loops that form at the temperature $T_{\rm {rh}}$ have a typical radius given by eq.~(\ref{lenloops}). This length is larger than $1/m_s$ provided that
\begin{equation}\label{looplungo}
\eta\gta 10^{15}\,{\rm {GeV}}\,\left(\frac{T_{\rm {rh}}}{10^9\,{\rm {GeV}}}\right)\,\left(\frac{{\rm {TeV}}}{m_s}\right)^{1/2}\,\,,
\end{equation}
where we have set $\nu=1,\,\left(\kappa-1\right)=10,\,\Gamma=100\,$. In order to simplify the discussion, we will assume that the above condition is satisfied.
The number of particles emitted by every loop will be given, neglecting the logarithmic factor in eq.~(\ref{n1}), by
\begin{equation}
N_s\sim \alpha^2\,\frac{\eta^2}{\Gamma\,m_s^2}\,\,.
\end{equation}
Hence, the number density of moduli produced in the time interval $t_F < t
< t_F + d t_F\,$, and redshifted to the time $t_* = M_P/T_*^2 > t_F\,$, amounts
to
\begin{equation}
d n_s \sim N_s\,\frac{d n_L}{d t} \left( t_F \right) d t_F \, \left( \frac{t_F}{t_*} \right)^{3/2} \sim \alpha^2 \, \frac{\nu}{\left(\kappa-1\right)\,\Gamma^2}\,\frac{M_P^2}{m_s^2}\,\frac{dt_F}{M_P^{3/2}\,t_F^{5/2}}\,T_*^3\,\,.
\label{dn}
\end{equation}
This last expression has then to be integrated over the time $t_F$ at
which loops form. It is clear that the earliest possible times $t_{\mathrm
{in}}$ dominate the integral. According to the above discussion, we set
$t_{\rm in}$ to be the time at which reheating ends. Therefore, dividing
the total number $n_s$ of moduli produced by the entropy $s\sim
T_*^3$,~\footnote{With the analogous computation, one can show that --
starting form eq.~(\ref{ew}) -- the energy density in quanta of $w$
produced by the loops and redshifted to the time $t_*$ amounts to
\begin{equation}
\rho_w \simeq \frac{g^2 \, \nu}{\lambda^{3/2} \left( \kappa -1 \right) \, \Gamma^2} \, \frac{M_P \, T_{\rm rh}^2}{\eta^3} \, T_*^4 \sim 10^{-14}\,\frac{g^2}{\lambda^{3/2}}\,\left(\frac{10^{15}\,{\rm {GeV}}}{\eta}\right)^3\,\left(\frac{T_{\rm {rh}}}{10^{9}\,{\rm {GeV}}}\right)^2\,T_*^4\,\,.
\end{equation}
Thus, in the application we are considering, we can neglect the increase of entropy due to the decay of the loops.} we get
\begin{equation}
Y_s\equiv \frac{n_s}{s} \sim \frac{\alpha^2\,\nu}{\left(\kappa-1\right) \, \Gamma^2} \, \frac{T_{\rm rh}^3}{m_s^2 \, M_P}\sim 10^{-2}\,\alpha^2\,\left(\frac{T_{\rm {rh}}}{10^9\,{\rm {GeV}}}\right)^3\,\left(\frac{{\rm {TeV}}}{m_s}\right)^2\, \,.
\label{finab}
\end{equation}

Therefore, we see that for the typical values $T_{\rm {rh}}\simeq
10^9\,{\rm {GeV}}$, $m_s\simeq {\rm {TeV}}$, $\eta \simeq 10^{15}\,{\rm
{GeV}}$, the coupling $\alpha$ is constrained to be smaller than about
$10^{-6}$ (whereas it could be expected to be of the order of unity) if we
do not want to violate the nucleosynthesis bound~(\ref{bound}).

\section{Bounds on the string scale}\label{scale}

As we have seen in the previous section, if the scale $\eta$ of the string
network is the natural scale of hybrid inflation, i.e. above about
$10^{15}$~GeV, a strong constraint is imposed either on the reheating
temperature or on the coupling $\alpha$. In the present section, on the other
hand, we will assume that the coupling $\alpha$ gets its ``natural'' value of
the order of unity and that $T_{\rm {rh}}\simeq 10^9\,$GeV, and we will
therefore derive a constraint on the scale $\eta\,$. An estimate in this sense
has been carried out in~\cite{davi}, where, for modulus mass of the order of
the TeV, the constraint $\eta \lta 10^{11}$~GeV was obtained. In obtaining this
result, the authors of ref.~\cite{davi} referred to a regime in which the
length of the string loops is much {\em{smaller}} than the Compton wavelength
of the modulus $\sim m_s^{-1}\,$. However, it may be possible that other decay
channels are relevant for such small loops, so that the time evolution of the
loop lenght is not given anymore by the expression~(\ref{radius}). This will in
turn affect the total number of moduli emitted by a single string loop during
its life, eventually modifying the bounds obtained in~\cite{davi}.

Indeed, a new decay channel that can be expected for the small loops is
given by cusp annihilation. As we have discussed in section~\ref{unaloop},
the description of a delta--like string breaks down when we get close to
cusps, where two portions of the strings overlap. As a consequence, the
cusp is generically expected to decay into a burst of quanta of the field
$\phi\,$, although the details of this process can be hardly described
analytically, and the power emitted by cusp annihilation can be only
roughly estimated \cite{brand,blpi}. In the following analysis we will assume the validity of the more conservative estimate given in~\cite{blpi}
\begin{equation}
P_c\sim \eta^2 \,\frac{1}{\sqrt{\eta\,L}}\,\,.
\end{equation}

When comparing $P_c$ with $P_g$, we see that the dominant decay channel for the string loop is given by cusp annihilation for loops whose length satisfies
\begin{equation}
L\lta  L_c=\frac{M_P^4}{\Gamma^2\,\eta^5}\,\,.
\end{equation}

For $m_s \sim \,$ TeV, the length $L_c$ is larger than $m_s^{-1}$ only if $\eta\lta 10^{15}\,$GeV. We thus see that the annihilation of the cusps does not affect the results presented in the previous section, where $\eta$ was taken to be slightly above this value. However, for smaller $\eta$, loops smaller than about $1/m_s$ will evolve as
\begin{equation}
\eta^2\,\frac{dL}{dt}\sim \eta^2 \,\frac{1}{\sqrt{\eta\,L}}\,\,.
\end{equation}
rather than according to eq.~(\ref{radius}). As a consequence, in this regime, the total number of moduli emitted by loops formed with a length $L_F$ will be given by 
\begin{eqnarray}\label{numcusps}
N_s\simeq\left\{\begin{array}{l}
\frac{\alpha^2}{M_P^2}\,\frac{\eta^{9/2}\,L_F^{1/2}}{m_s^2}\,\,,\quad m_s^{-1}\lta L_F\lta L_c\,\,,\\
\frac{\alpha^2}{M_P^2}\,\eta^{9/2}\,L_F^{5/2}\,\,,\quad L_F\lta m_s^{-1}\lta L_c\,\,,\end{array}\right.
\end{eqnarray}

Using the above formulae, it is possible to analyze systematically the
production of moduli that can have been occurred during the various
cosmological eras. As in the previous one, also in the present section we
will neglect particle production that could have been occurred during
reheating and during the friction dominated era (notice that the same
conservative hypothesis is made in~\cite{davi}). In this case, the
analysis of the total number of moduli emitted by the string network shows
that they are mainly produced when the typical length of loops formed by
the network is of the order of $m_s^{-1}\,$, that is when the temperature of the Universe is about $T \sim 10^{-7}\, \eta$ (for the usual values $m_s \sim\,$ TeV, $\left( \kappa - 1 \right) \sim 10\,$, $\Gamma \sim 100\,$). Then, the abundance of moduli produced for $\eta\lta 10^{14}\,$GeV turns then out to be given by
\begin{equation}
Y_s\sim 10^{-5}\,\alpha^2\,\left(\frac{\eta}{10^{15}\,{\rm {GeV}}}\right)^{11/2}\,\,.
\end{equation}
This result, for $\alpha\sim 1$, implies a bound 
\begin{equation}
\eta\lta 10^{13}\,{\rm{GeV}}
\end{equation}
for a cosmic string network that is coupled via $1/M_P$ coupling to scalar moduli of mass of the order of the TeV. Notice that this bound is actually weaker by a couple of orders of magnitude than the analogous one presented in ref.~\cite{davi}. The difference can be simply accounted for by the total number of moduli emitted by a string loop: if cusp annihilation effects are neglected, one obtains that, for loops of length $\sim m_s^{-1}$, $N_s \sim \alpha^2\,\eta^2/m_s^2$~\cite{davi}, whereas, if cusp annihilation is assumed to be effective, eq.~(\ref{numcusps}) gives $N_s\sim \alpha^2\, \eta^{9/2} / (M_P^2\,m_s^{5/2})$ that, for the values of $\eta$ we are interested in, leads to a much less efficient production of moduli.

\section{Conclusions}\label{concl}

We have considered the production of scalar moduli by a network of cosmic
strings. In particular, we have analyzed the simple case of local,
nonsuperconducting strings that obey the Nambu-Goto dynamics. We
considered only the production that took place in cosmological epochs in
which the interaction with the background plasma did not significantly
affect loop dynamics, and during which we disposed of a reliable estimate
of the number density of loops present in the Universe. Despite the above
conservative assumptions, the production can be rather efficient. In
particular, a cosmic string network formed at the end of hybrid inflation
at a scale of about $10^{15}\,$GeV can lead to a moduli abundance that
exceeds by several orders of magnitude the Big Bang Nucleosynthesis bound.
Such conclusions can be avoided either by assuming a low reheating
temperature ($T_{\rm rh} \lta 10^6$ GeV), or by requiring that the string
coupling to moduli is much smaller than the typical gravitational coupling
$1/M_P\,$. This second option is certainly a possibility, since the
coupling of moduli to the fields which form the strings is a model
dependent quantity, although it is hard to expect such a small coupling to
arise naturally. Actually, if we assume $\alpha\sim 1$ (and take $T_{\rm rh} \simeq 10^9 \,$ GeV), the energy scale of the string has to be smaller than about $10^{13}$~GeV not to conflict with Big Bang Nucleosynthesis. This bound is of few orders of magnitude weaker than the one reported in the previous literature~\cite{davi}, due to the fact that cusp annihilation effects can lead to a faster decay of loops of strings, thus decreasing the amount of moduli produced.

In our computation, we have made an important distinction between processes
characterized by the production of a single modulus (from a trilinear vertex
$\phi^2 \, s$) and processes with the production of two quanta (from quartic
vertices $\phi^2 s^2$). In the former case, the production is mainly due to
collective oscillations of the whole loop, while in the latter high frequency
and small scale oscillations dominate. In this second case, the computation may
be affected by nonperturbative phenomena leading to cusp annihilation which
occur when high frequency oscillations lead different portions of the loop to
be closer to each other than the width of the loop itself. The bounds we have
derived come from single quanta emissions, which are insensitive to these small
scale effects.

As we have stressed in the Introduction, moduli fields are not the only
gravitational relics that one should be worried about. In particular, as
was also noted in~\cite{kklv}, production of gravitinos by the oscillating
loops may turn out very efficient, since in this case we know the coupling
to be of gravitational strength. The production of gravitinos is
technically more involved than the one for scalar particles presented
here. Moreover, being a two particles production, nonperturbative effects
may influence the final result.

\section{Acknowledgments}

It is a pleasure to thank Antonio Riotto for bringing our attention to the
problem and for interesting comments, and Gianmassimo Tasinato for
collaboration at the early stages of the work. We acknowledge useful
conversations with Robert Brandenberger, Stefan Groot Nibbelink, Lev Kofman,
Patrick Peter, and Daniele Steer. We also thank Pierre Bin\'etruy, Jose
J.~Blanco-Pillado, and Anne Davis for important comments on the
earlier version of the paper. This work is partially supported by the European
Community's Human Potential Programme under contract HPRN-CT-2000-00152
Supersymmetry and the Early Universe. M.P. also acknowledges support by the
contracts HPRN-CT-2000-00131 and HPRN-CT-2000-00148.

\end{document}